# High-yield exfoliation of MoS$_2$ nanosheets by a novel spray technique and the importance of soaking and surfactants

Suvigya Kaushik[1#], Siva Sankar Nemala[1#], Mukesh Kumar[2], Devesh Negi[3], Biswabhusan Dhal[1], Lalita Saini[1], Ramu Banavath[4], Surajit Saha[3], Sudhanshu Sharma[2] & Gopinadhan Kalon[1,5] *

[1]Discipline of Physics, Indian Institute of Technology Gandhinagar, Gujarat 382355, India

[2]Discipline of Chemistry, Indian Institute of Technology Gandhinagar, Gujarat 382355, India

[3]Department of Physics, Indian Institute of Science Education and Research Bhopal, Bhopal, 462066, India

[4]Department of Metallurgical Engineering and Materials Science, Indian Institute of Technology-Bombay, Mumbai 400076, India

[5]Discipline of Materials Engineering, Indian Institute of Technology Gandhinagar, Gujarat 382355, India

(*E-mail: gopinadhan.kalon@iitgn.ac.in)

(#-Equally contributed)

**Liquid-phase exfoliation of two-dimensional materials is very attractive for large-scale applications. Although used extensively, isolating MoS$_2$ layers (<10) with high efficiency is reported to be extremely difficult. Further, the importance of soaking has not yet been studied, and the surfactants' role in stabilizing MoS$_2$ nanosheets is poorly understood[1]. Herein, we report a novel approach to exfoliating large quantities of MoS$_2$ via high-pressure (HP) liquid-phase exfoliation (LPE) in deionized (DI) water. 4 to 7 layers of MoS$_2$ nanosheets were obtained from 60 days-soaked samples and they were found to be stable in solvents for periods of up to six months. Studies on the effect of three surfactants, namely sodium dodecyl benzenesulfonate (SDBS), sodium cholate (SC), and tetra-butyl ammonium bromide (TBAB), indicate that exfoliation of MoS$_2$ nanosheets in SDBS is highly efficient than the other two surfactants. The estimated yield reaches up to 7.25%, with a nanosheet concentration of 1.45 mg/ml, which is one of the highest ever reported. Our studies also suggest that the nanosheets' concentration and the lateral size depend on exfoliation cycles, applied pressure and surfactant concentration. Hydrogen evolution reaction (HER) and ion-transport study show that the nanosheets prepared by our method are stable in an acidic medium and free from surfactants. A high hydrogen evolution rate of 30.13 mmol g$^{-1}$ h$^{-1}$ was estimated under ambient laboratory conditions.**

1. Introduction

Transition metal dichalcogenides (TMDs) have been increasingly explored due to several useful characteristics, such as atomically small thickness and large spin-orbit coupling, which promises a wide range of applications[2]. The TMDs are represented by the chemical formula $MX_2$, where M is a transition metal (Mo, W, etc.), X is a chalcogen (S, Se, Te, etc.), and they have a hexagonal structure. In $MX_2$, one M layer is sandwiched between two X layers with strong covalent bonds within one layer but weak van der Waals interaction between adjacent layers. This makes it feasible to exfoliate into mono- or few-layered nanosheets.

Among TMDs, $MoS_2$ crystals are abundant, which prompted us to focus on $MoS_2$ nanosheets. $MoS_2$ sheets can be arranged into nanocapillary channels exhibiting ion sieving effects, thus useful for desalination and nanofiltration applications[3]. Due to its layered structure, $MoS_2$ is an excellent lubricant in space applications[4]. Monolayer $MoS_2$ is a direct bandgap semiconductor ( ~1.8 eV) in contrast to bulk (~1.23 eV), making it useful for transistors[5] and solar energy harvesting[6]. Few-layered $MoS_2$ is equally interesting for applications such as flexible and wearable electronics[7], sensors[8–11], drug delivery[12], supercapacitor[13], and photo-catalysis[14]. $MoS_2$ nanosheets have more surface-active sites than the bulk, which is expected to enhance its electrocatalytic activity. These nanosheets have a high surface area, thermal stability, electrical conductivity, and adsorption capability[15]. Several studies indicate their potential for immediate applications as a catalyst in HER[15–22].

The large-scale synthesis of $MoS_2$ has been attempted several times, but it remains a challenge to produce good quality and large quantities commercially[5,14,23,24]. The commonly used exfoliation method of ultra-sonication can accommodate only a limited volume of parent dispersion. High-shear methods[25,26] employ rotor-stator shear mixers or kitchen blenders. They have been reported to produce nanosheets similar in quality and size to those obtained from sonication, with high production rates. Although the high-shear method is better than sonication in volume, it is inefficient. A huge high shear mixer can compensate for the volume but leaves a large amount of material (away from the rotor blade) unexfoliated. It also requires high power for operation. The attempts in this field are directed towards increasing the exfoliation efficiency but have not been as successful as exploiting the best out of the technique of liquid-phase exfoliation. Wang *et al.* exfoliated using sonication and achieved the final nanosheets concentration of 0.2 mg/ml[27]. Using a solvothermal treatment with formamide, Huang *et al.* found the concentration of the $MoS_2$ nanosheets to be 0.21 mg/ml[28]. Liu *et al.* achieved a 0.24 mg/ml by a high-pressure method, wherein they improved the exfoliation efficiency of $MoS_2$ nanosheets in isopropanol using inexpensive salts[23]. The high shear mixing technique applied by Varrla *et al.* led to a better concentration of 0.4 mg/ml[25]. Recently, Janica *et al.* exfoliated $MoS_2$ nanosheets using lithium compound and achieved 0.5 mg/ml concentration[12]. Hai *et al.* employed a surfactant-free method to increase $MoS_2$ monolayers' yield and achieved a concentration of 0.45 mg/ml in isopropanol[29]. Earlier, Lin *et al.* produced $MoS_2$ nanosheets with a yield of 52% in $NaNO_3$/HCl solution, but the method utilized hazardous acids[30].

In this paper, we explore an eco-friendly and safe high-pressure liquid-phase exfoliation approach to overcome issues associated with large-scale and high-yield production of stable $MoS_2$ nanosheets. A large quantity of $MoS_2$ (>1000 ml) was exfoliated with this method using deionized (DI) water as the solvent, which has been deemed more complicated than exfoliation in NMP[28] or DMF. This is due to the difference in surface energy of DI water (72.2 mJ m$^{-2}$) and $MoS_2$ (46.5 mJ m$^{-2}$)[31], whereas the surface energy of NMP is 40 mJ m$^{-2}$ [32]. Molybdenum disulfide, being difficult to exfoliate with other techniques, has been successfully exfoliated and characterized in this paper. We achieved 7.25% exfoliation efficiency and 1.45 mg/ml nanosheets concentration, considerably higher than the previous methods.

## 2. Experimental section

A. Liquid-phase exfoliation of MoS$_2$

MoS$_2$ exfoliation was done using a novel airless spray exfoliation technique. The details of materials used for the exfoliation are provided in the Supporting Information (Section 1). This method consisted of several steps: intercalation, expansion, exfoliation, and separation. In the initial phase, 1 mg/ml of SDBS was added to water and stirred for 15 minutes at room temperature to get a transparent solution. Then, 20 mg/ml of MoS$_2$ bulk powder was taken in a large container (1.5 L), and water was added to it, to which the prepared SDBS transparent solution was mixed. The solution was sonicated for 30 minutes to get a homogeneous solution. The prepared solution was kept for soaking for a long duration of 15 & 60 days at room temperature under dark conditions for the proper intercalation. This process weakened the van der Waals force between the individual MoS$_2$ layers. It is expected that the expanded MoS$_2$ flakes can then be easily exfoliated into nanosheets at low pressure/force.

Finally, the prepared intercalated MoS$_2$ solution was passed through the nozzle of an airless paint sprayer (AEROPRO Painter R450, Flow rate 2.2 L min$^{-1}$) (Fig. 1 (a)) and collected in another container (Fig. 1 (b)). The shear force created at the nozzle (Fig. 1 (a)) helped to separate individual MoS$_2$ nanosheets. The collected solution was again passed through the nozzle, repeated several times, and hereafter referred to as cycles. The final solution was collected and centrifuged at 5000 rotations per minute (rpm) for 30 minutes to separate the smaller and larger flakes. The finer MoS$_2$ nanosheets were collected from the top two-thirds of the centrifuge tube. Surfactant removal is a crucial step and is not much explained in the literature[1]. We tried to remove the surfactant to obtain pure nanosheets. The nanosheets, after centrifugation, were filtered on PVDF (poly (vinylidene fluoride)) or AAO (anodic aluminium oxide) substrate (pore size 0.1 µm) and washed first with acetone and then isopropanol. Without letting them dry, the nanosheets were scooped from the surface of AAO using a spatula and dispersed in isopropanol again for further characterizations. We found this method very efficient for obtaining pure nanosheets devoid of surfactant.

B. Electrochemical Studies

The electrochemical analysis of MoS$_2$ nanosheets was done using a three-electrode system connected to Metrohm Autolab electrochemical workstation with NOVA software. Platinum (Pt) wire was used as the counter electrode (CE), Ag/AgCl as the reference electrode (RE), and glassy carbon electrode (GCE) as the working electrode (WE). The surface of GCE was polished with alumina powder and cleaned with DI water several times before the preparation of the electrode. A homogeneous dispersion of the MoS$_2$ nanosheets (slurry) was prepared in Nafion (1:1 w/w) and isopropanol and ultrasonicated for 30 minutes. 3 µl of this slurry was then drop-casted on GCE (diameter = 3 mm, surface area = 0.071 cm$^2$) to prepare the WE. 0.5 M sulphuric acid (H$_2$SO$_4$) served as the electrolyte.

Linear sweep voltammetry (LSV) was performed at a slow scan rate of 5 mV s$^{-1}$ in the range of 0 V to -0.8 V and used to determine the Tafel slopes. Cyclic voltammetry (CV) of MoS$_2$ nanosheets was performed at a narrow voltage window (from 0 V to -0.1 V) at different scan rates (20-100 mV s$^{-1}$) to obtain the double layer capacitance. To check the stability of the nanosheets, chronoamperometry (i-t) studies were conducted for 24 hours under constant stirring. Prior to all the studies, the LSV/CV response of the blank GCE (without the nanosheets coating) was recorded. Then, the GCE was coated with nanosheets and inserted into the system. Nitrogen gas was purged for 20 minutes to create an inert atmosphere and eliminate the dissolved oxygen. Gas chromatography was conducted to investigate the gaseous products formed during the reaction. The system was sealed with parafilm before the experiment. The gaseous products from the sealed system were collected in an air-tight syringe (Hamilton, 2ml) and analyzed with a gas chromatography instrument (CIC Dhruva, Baroda). It is equipped with a thermal conductivity detector (TCD) for identifying hydrogen, which was calibrated using a standard gas mixture. The number of moles of hydrogen produced by 1 ml of collected gaseous products was calculated from the area under the peak of the gas chromatogram.

C. Ion-transport Study

A 9 mm x 3 mm rectangular strip was cut from an $MoS_2$ membrane prepared on PVDF (pore size 0.1 µm). It was encapsulated with the PDMS in such a way that the only transport path for ions was through the planes of the membranes. This was then heated at 100 $^0$C for 10 minutes in the convection oven till the PDMS was set. It was ensured that out-of-plane transport was prohibited by compact encapsulation. The thickness of the PDMS was ~ 3 mm.

The voltage was varied from -200 mV to +200 mV using a Keithley 2614B source meter. The resulting current was measured using Ag/AgCl electrodes through a LabVIEW program connected with the source meter. Aqueous KCl solutions with concentrations from 10$^{-5}$ M to 1 M were used as electrolytes.

**3. Results and Discussion**

The dispersion of the nanosheets can be probed by a laser beam. The path of the laser beam is visible through the $MoS_2$ solution (Fig. 1 (c)), which indicates that the sample is a colloidal dispersion of $MoS_2$ nanosheets; this phenomenon is called Tyndall effect. Different concentrations of $MoS_2$ nanosheets prepared by varying the exfoliation parameters are shown in Fig. 1 (d).

The Supporting Information (Section S2) describes the details of the techniques used to characterize the exfoliated nanosheets. The atomic force microscopy (AFM) image in Fig. 2 (a) with the height profile provides a flake thickness of 4-5 nm. Based on the interlayer spacing of 0.67 nm for $MoS_2$, our exfoliated sheets primarily consist of 6-7 layers. Transmission electron microscopy (TEM) analysis of $MoS_2$ was done to confirm its exfoliation and the size of nanosheets. Fig. S1 (a) shows the high-resolution TEM (HR-TEM) image of exfoliated $MoS_2$ flakes confirming that the high-pressure exfoliation process is efficient in exfoliating bulk $MoS_2$ to few layers. These nanosheets are a few hundred nanometers in size. The fringe spacing of 0.27 nm is calculated from the HR-TEM image (Fig. S1 (b)). As reported previously[33], it corresponds to the (100) plane of $MoS_2$, which is shown as the inset of Fig. 2 (b). A Moiré pattern is clearly visible in our TEM images (Fig. 2 (b)), very similar to the previously reported works[31,32]. These fringes arise in few-layered $MoS_2$[31] or from the boundary between two grains[32]. The appearance of the pattern is due to the rotation of the top flake by a slight angle with respect to the bottom flake. It is further evidence of the thinness of our $MoS_2$ nanosheets.

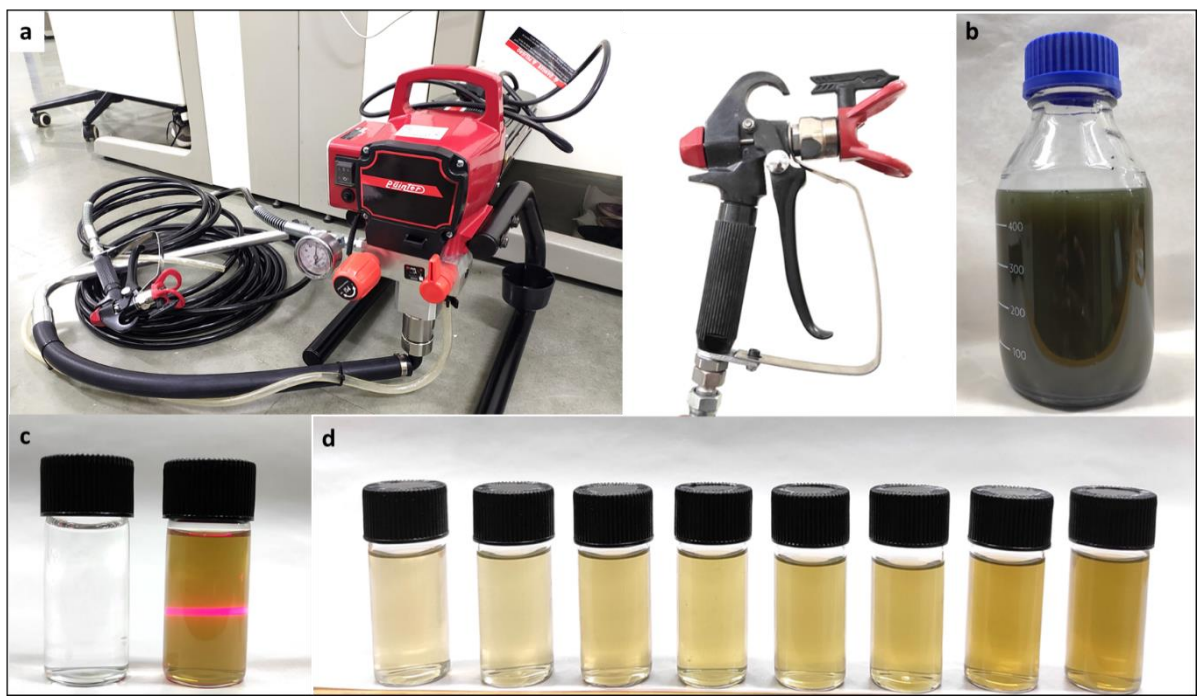

**Fig. 1: Liquid-phase exfoliation of MoS₂ using high-pressure spray technique.** (a) The airless paint sprayer and the side view of the spray gun with nozzle used for the high-pressure liquid phase exfoliation. (b) A large amount of MoS₂ solution collected after exfoliation and centrifugation and stored in a glass bottle. (c) Tyndall effect in the MoS₂ nanosheets shows that the path of the red laser is visible through the colloidal suspension of nanosheets (water on the left side shows no such property). (d) Different concentrations of MoS₂ nanosheets prepared by varying exfoliation cycles and applied pressure.

Raman spectroscopy is a non-destructive characterization technique that provides structural and electronic information about materials. The Raman spectra of MoS₂ nanosheets and bulk MoS₂ are shown in Fig. 2 (c). The two sets of peaks correspond to $E_{2g}^1$ and $A_{1g}$ modes, indicating the in-plane and out-of-plane vibrational modes of the MoS₂ layer[34]. The peak position and the frequency difference (the peak-to-peak separation) of these two modes are sensitive to the layer thickness of MoS₂. In bulk MoS₂, the peak-to-peak separation of these modes is 25.2 cm$^{-1}$. After exfoliation, $A_{1g}$ and $E_{2g}^1$ modes shifted towards the higher wavenumber. The peak-to-peak separation is estimated to be 24.6 cm$^{-1}$ for MoS₂ sheets that are exfoliated after 60 days of soaking. This indicates that four or fewer layers are present[23] in the exfoliated MoS₂. Moreover, the shift of $A_{1g}$ mode is more remarkable than that of the $E_{2g}^1$, which indicates that the number of layers has been reduced to a few layers of MoS₂ by weakening the van der Waal forces between the layers during the exfoliation process. These results verified the successful exfoliation of the material via the novel high-pressure liquid-phase exfoliation technique and the existence of nanosheets. Fig. S1 (c) shows that the Raman shift for the 15 days nanosheets is more than that for 60 days, suggesting that layer number reduces significantly upon increasing the soaking duration.

X-ray diffraction (XRD) was performed to understand the crystalline nature of the exfoliated material. Fig. 2 (d) shows the XRD patterns of bulk $MoS_2$ and the 60 days nanosheets on AAO. The XRD pattern of bulk $MoS_2$ shows an intense peak at 14.86°, which corresponds to the diffraction from the (002) plane, indicating that the bulk material consisted of a large number of layers oriented in that particular direction. Similarly, the prepared $MoS_2$ nanosheets exhibit the exact peak of the (002) plane with reduced intensity and weak broadening, implying that bulk material was exfoliated into few-layered nanosheets upon exposure to the high-pressure[35]. Again, the characteristic peak shows a small shift towards the lower angles of 2θ, which confirms that the interlayer spacing along the c-axis is affected; this may result from surfactant intercalation between the layers of $MoS_2$ during the process. No secondary phases are detectable in the XRD data, indicating that the flakes are not oxidized during the process. The average crystallize size estimated using Scherrer's formula[36] is approximately 11 nm (Section S5.1). The XRD pattern for 15 days nanosheets is shown in Fig. S1 (d).

The field-emission scanning electron microscopy (FE-SEM) image and the corresponding electron dispersive X-ray spectroscopy (EDS) were also recorded to estimate the lateral size and elemental analysis of the exfoliated flake of $MoS_2$, respectively Fig. S2 shows the FE-SEM images of $MoS_2$ nanosheets, along with EDS results. The FE-SEM image (Fig. S2 (a)) provides a flake size of a few

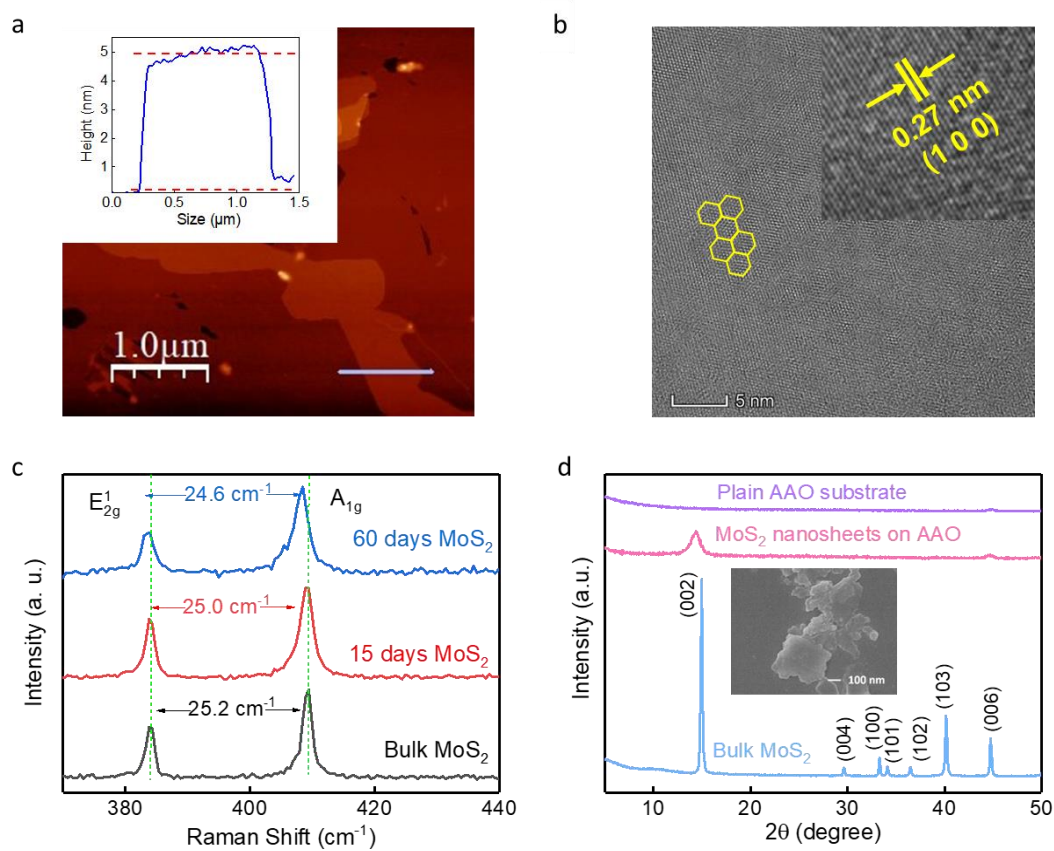

**Fig. 2: Characterizations of exfoliated $MoS_2$ nanosheets:** (a) AFM image of the nanosheets collected by vacuum filtration and the corresponding height profile. (b) HR-TEM image of $MoS_2$ nanosheets shows Moiré pattern. The thin layers of $MoS_2$ in hexagonal symmetry are on top of each other and rotated by a small angle with respect to the bottommost layer; the inset shows an interlayer spacing of 0.27 nm, which corresponds to the (100) plane. (c) Raman spectra for bulk $MoS_2$ and $MoS_2$

nanosheets soaked for 15 and 60 days; the reduced peak-to-peak separation indicates a reduction in the number of layers upon increasing the soaking duration. (d) XRD pattern of the 60 days nanosheets compared with the bulk (JCPDS No.: 06-0097) and bare substrate AAO; the SEM image in the inset shows the size of the exfoliated nanosheets, which is a few hundred nanometers.

hundred nanometers, which agrees with the result of TEM. The corresponding EDS (Fig. S2 (b)) results reveal the purity of exfoliated $MoS_2$ nanosheets. The elemental composition of exfoliated $MoS_2$ sample matches the elemental composition of pristine $MoS_2$, which confirms the absence of any modification in the chemical nature of exfoliated $MoS_2$ nanosheets. The absence of elements other than Mo and S in the EDS spectra proved that left-out additives and intercalating agents were removed entirely from exfoliated nanosheets through washing with acetone and isopropanol.

Ultraviolet-visible (UV–visible) absorption spectrum of $MoS_2$ soaked in different surfactants for 60 days is shown in Fig. 3 (a). The spectra have two well-defined absorption peaks at 610 nm and 670 nm, which are the characteristic peaks of $MoS_2$ nanosheets. These originate from B and A direct excitonic transitions in the first Brillouin zone occurring at the K/K' points[23]. A weak and broad peak was observed at 450 nm; it could be arising from C excitonic transition, a transition between the deep valence band and the conduction band. This result implies that a large number of exfoliated $MoS_2$ nanosheets exist in the prepared dispersion. We found that the exfoliation was better in SDBS than in the other two surfactants as evident from the presence of intense excitonic peaks in the case of SDBS alone. Fig. S3 (a) shows the spectra of prepared $MoS_2$ nanosheets in SDBS surfactant soaked for 60 days and 15 days. It is evident that those soaked for a more extended period showed more intense peaks. We used the UV-Visible spectra to estimate the band gap of the nanosheets, which is ~ 1.8 eV for both 60 days and 15 days soaked samples (Fig. S3 (b)), similar to the values reported in the literature[5,14].

We calculated the zeta potential of the solution with SDBS as the surfactant, which is ~-33 mV (Fig. S3 (c)), indicating that the prepared nanosheets are stable[37]. 10 ml of exfoliated $MoS_2$ was centrifuged at 1500, 2000, and 3000 rpm for 15 minutes and kept in a lyophilizer for 72 hours. The concentration of nanosheets was found to be 1.70, 1.68, and 1.56 mg/ml, respectively. Yield as high as 8.5% was achieved for the sample centrifuged at 1500 rpm. At 5000 rpm, very fine nanosheets were obtained, and the concentration was 1.45 mg/ml, still giving a yield of 7.25%. However, the freeze-dried nanosheets might contain traces of the surfactant and thus affect the concentration estimation. Therefore, the concentration of $MoS_2$ nanosheets was re-examined using Lambert-Beer law, $A/l=\alpha C$, where A is the absorbance, l is cell length, $\alpha$ is the extinction coefficient, and C is the concentration (Fig. 3 (b-d)). It resulted in a nanosheets' concentration of 0.96 mg/ml (Section S5.2).

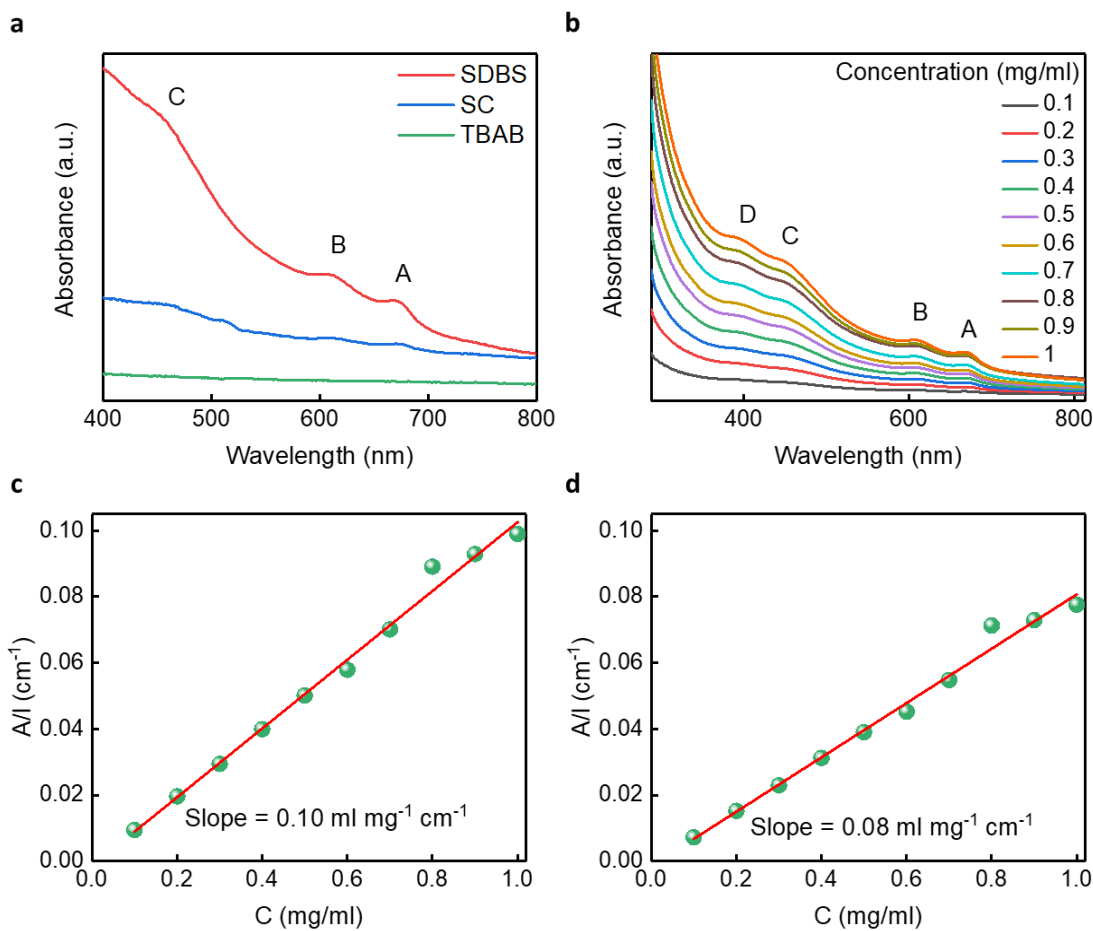

**Fig. 3: UV-Visible absorption spectra:** (a) Absorbance of solutions of MoS$_2$ soaked in different surfactants SDBS, SC, and TBAB for 60 days. (b) Absorbance for different concentrations of nanosheets soaked in SDBS for 60 days (here, the concentration of the master solution is termed 1, and all other concentrations are prepared by its dilution). Absorbance per unit length vs. concentration at the two excitonic peaks B and A at (c) 610 nm and (d) at 670 nm, respectively, to determine the value of extinction coefficient (Section S5.2).

In a recent report, Coleman and colleagues discussed the effect of the choice of surfactant and its concentration on the yield of the nanosheets[38]. The authors chose various ionic and non-ionic surfactants and found that "surfactant choice is less important than might be expected." We tried to look deeper into this statement through our work, hoping to see similar results in the case of all the three surfactants used- SDBS, SC, TBAB. We prepared the solutions of MoS$_2$ using these surfactants in the same manner, exfoliated, and centrifuged with the same parameters as earlier. We instead found that the MoS$_2$ nanosheets in the case of SC and TBAB gave significantly less concentration and poor yield. It is also very evident from the UV-Visible studies (Fig. 3 (a)). The stability of MoS$_2$ nanosheets was, in fact, too poor in TBAB, and they settled at the bottom of the centrifuge tube quickly.

The control study was conducted with MoS$_2$ nanosheets (a similar concentration solution soaked for 15 days before exfoliation). Raman shift (Fig. 2 (c), Fig. S1 (c)) indicates that the nanosheets can be reduced to fewer layers with increasing soaking duration. The data from the UV-Visible spectroscopy (Fig. S3 (a)) also suggests that increasing the period of soaking of the $MoS_2$ in the surfactant increases the yield, which was 1.45 mg/ml when soaked for 60 days as discussed earlier, in contrast to 0.36 mg/ml for 15 days soaking. This emphasizes the role of soaking, which eases the layer separation process during the high-pressure exfoliation.

To enhance and tune the exfoliation process[39], it is quite necessary to study the influence of some critical parameters. The effect of variation of exfoliation cycles, applied pressure, and surfactant concentration was studied in detail and provided in Supporting Information (Section S4). The concentration increases with increase in the number of cycles at all applied pressures (Fig. S4 (a)). An increase in applied pressure reduces the lateral size, and it dominates more than the cycle number (Fig. S4 (b-d)). Further, an increase in the surfactant concentration reduced the concentration of the nanosheets. If the surfactant concentration remains smaller than 3 mg/ml, the nanosheet concentration shows tunability with number of cycles and applied pressure, whereas cycles and pressure influence very little at higher surfactant concentrations (Fig. S4 (e-f)).

A reduction in the nozzle size (here, 279.4 μm) could possibly result in more fragmentation of the flakes, making flake size even less. The smaller the flake size, the more difficult it is to produce inks for inkjet printing of electronic circuits[7,40]. On the other hand, optimum flake size facilitates more accessible flake-to-flake transport and helps in the easy spreading of the ink[7]. We have already studied the effect of variation in the number of cycles, applied pressure, and surfactant concentration on nanosheet concentration. Increasing the initial concentration of $MoS_2$ could result in a higher concentration of the final nanosheets.

Electrochemical studies were conducted to test the $MoS_2$ nanosheets for hydrogen evolution[16,17]. A typical three-electrode setup was used for this purpose[18], and its schematic is given in Fig. S5. Fig. 4 (a) shows the LSV curves of bulk $MoS_2$ and nanosheets, from a potential range of 0.0 V to -0.8 V in 0.5 M $H_2SO_4$. A higher current density at -0.8 V is observed for the $MoS_2$ nanosheets obtained after 60 days of soaking than those after 15 days. This could be because of the fewer layers in the former compared to the latter. Tafel slopes were obtained from the linear region of the LSV curves and are shown in Fig. 4 (b). Pt has the lowest Tafel slope of 46 mV dec$^{-1}$. Our 60 days nanosheets exhibited higher HER performance, characterized by a Tafel slope of 189 mV dec$^{-1}$, than the 15 days nanosheets (223 mV dec$^{-1}$). The overpotentials at a current density of -10 mA cm$^{-2}$ were found to be -0.83 V and -1.0 V for 60 and 15 days nanosheets, respectively. The soaking duration, thus, played a vital role in enhancing the performance of the 60 days nanosheets.

Chronoamperometry was performed at -0.8 V overpotential in an acidic medium to check the long-term functional durability of these nanosheets. They were found to be stable for the measurement duration of 24 hours (Fig. 4 (c)). Only in the case of 60 days nanosheets do we observe constant bubbles at the working electrode during the chronoamperometry measurement (Fig. S6 (a)). This phenomenon is also evident from its chronoamperometry curve in Fig. 4 (c), where noise due to bubbling can be seen. The inset of Fig. 4 (c) is an enlarged view of the current for two hours of the study. The hydrogen peak in the gas chromatography data (Fig. S6 (b)) confirmed the evolution of hydrogen gas at the working electrode. The rate of evolution of $H_2$ gas was estimated to be 30.13 mmol g$^{-1}$ h$^{-1}$ (Section S5.3) under ambient conditions. Our experiments were done in the normal day-light conditions and we could not find any literature that reports $H_2$ concentration similar to our experimental conditions. Therefore, we compared our results against few reports that describe the photocatalytic $H_2$ evolution. In a recent report by Chen et al.[41], the photocatalytic HER rate for CIZS/$MoS_2$/CDs reached only up to 3.706 mmol g$^{-1}$ h$^{-1}$ even though it was higher than their $MoS_2$ (0.025 mmol g$^{-1}$ h$^{-1}$). A hierarchical heterostructure of $NiSe_2$/$MoS_2$ composite reported by Jia et al.[42] exhibited a HER rate of 2473.7 μmol g$^{-1}$ h$^{-1}$ under light illumination. Liang et al.[43] prepared O-1T/2H–$MoS_2$/p-g-$C_3N_4$ hybrids which displayed photocatalytic HER rate of 1487 μmol g$^{-1}$ h$^{-1}$. Our 60 days nanosheets displayed an HER rate which is considerably higher than most of the photocatalytic HER rates previously reported in literature.

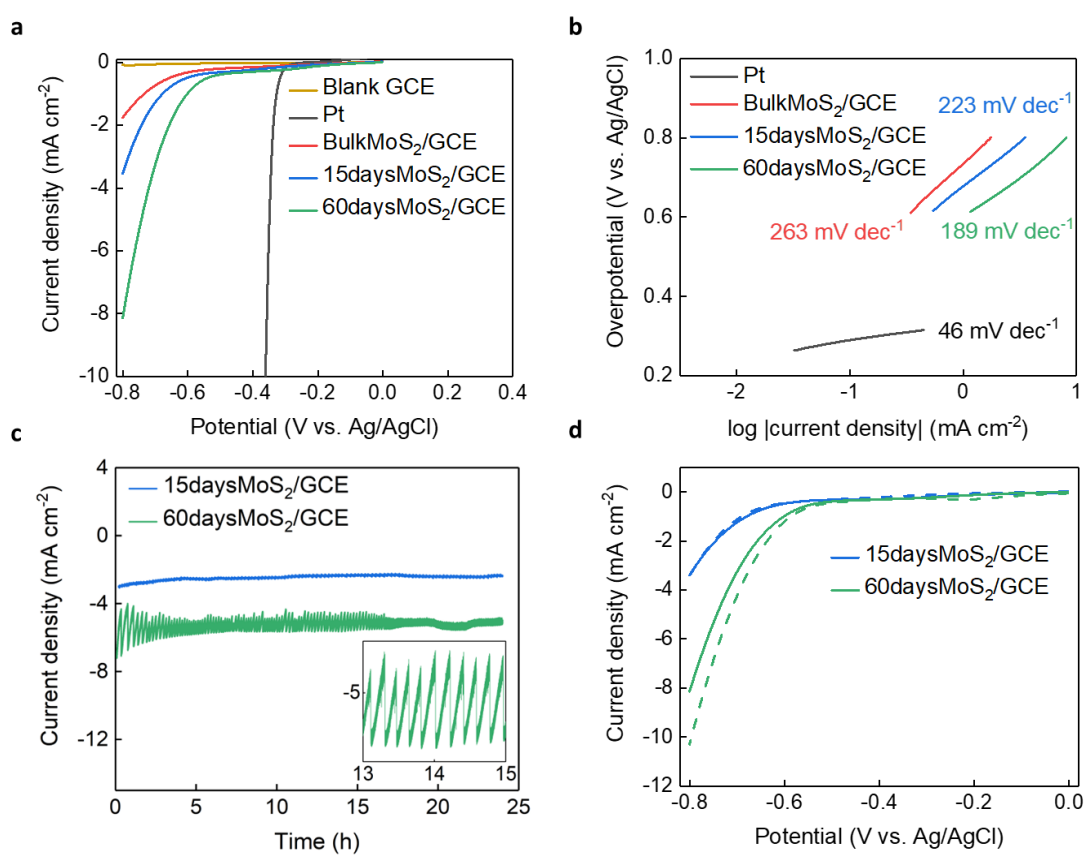

**Fig. 4: Electrochemical measurements:** (a) LSV curves of bulk $MoS_2$, 15 and 60 days nanosheets compared with Pt. (b) Corresponding Tafel slopes obtained from the LSV curves in (a). Stability tests: (c) Chronoamperometric (i-t) curves for 15 and 60 days nanosheets for 24 hours showing the stability of the nanosheets; the inset is an enlarged view of the current density variation between 13-15 hours of the study for 60 days nanosheets, representing the formation and release of the hydrogen gas bubbles, (d) LSV curves of 15 days and 60 days nanosheets before (solid lines) and after (dashed lines) the chronoamperometric studies also show the stability of the nanosheets in the acidic medium.

The stability of the nanosheets was also verified from the LSV curves obtained before and after 24 hours of the chronoamperometry study. As shown in Fig. 4 (d), the current densities were found to be similar for 15 days nanosheets demonstrating the stability of the material; however, the current increased marginally after the chronoamperometry for 60 days nanosheets.

The bulk $MoS_2$ had a higher Tafel slope of 263 mV dec$^{-1}$ than the nanosheets and an overpotential of 1.4 V at a current density of -10 mA cm$^{-2}$. The activity of the nanosheets is sufficiently higher because of the more edge sites available for catalytic activity than the bulk.

The capacitive currents were measured as a function of scan rates, as shown in Fig. S7 (a-b). The capacitance of the double layer ($C_{dl}$) at the solid-liquid interface was estimated with CV curves at a narrow voltage window of 0 V to -0.1 V at scan rates ranging from 20-100 mV s$^{-1}$. The $C_{dl}$ (Fig. S7 (c)) was found to be greater (0.33 mF cm$^{-2}$) for the 60 days nanosheets. This implies that their surface area is higher[19] than the 15 days nanosheets (0.18 mF cm$^{-2}$). The extended soaking thus enhanced the surface area of the 60 days nanosheets, resulting in a higher catalytic current.

Our nanosheets are found to be active and stable in HER application, with the obvious effect of an extended soaking period. However, it should be noted that their activity must be enhanced to reach the literature values of current densities and overpotentials[16]. Previous reports on hydrogen evolution reaction catalysis of two-dimensional $MoS_2$ state that the activity of the nanosheets can be further improved by doping, introducing defects, enhancing the per site activity, or making the inert basal plane active for catalysis[20,22]. Defect-rich nanosheets can significantly improve the catalytic activity in the nanosheets[21].

We conducted ion-transport study to confirm that the nanosheets are indeed surfactant-free. $MoS_2$ membrane was prepared on a PVDF substrate of pore size 0.1 μm by vacuum filtration (Fig. S8(a)). Using PDMS, we prepared a geometry allowing only the in-plane transport of ions. This sample was exposed to the electrolyte solution in an electrochemical cell with two reservoirs (Fig. S8 (b)) similar to that used in our previous work[44]. The prepared sample geometry is shown in Fig. S8 (b) and inset of Fig. S8 (c). We recorded the current by applying a potential across the membrane. There is no appreciable current up to the maximum applied voltage of 200 mV, shown in Fig. S8 (c). A similar current value is measured from the setup in the absence of any membrane, which suggests that the measured small current in the presence of a membrane is related to the leakage of the system. The obtained current of ~$10^{-11}$ A at an applied voltage of ±200 mV remains the same for all the concentrations (Fig. S8 (c)). The interlayer space of $MoS_2$ does not allow the hydrated ions to pass through it, thus giving only a leakage current. The ion transport results confirm that the $MoS_2$ nanosheets are pure and surfactant-free.

This novel high-pressure LPE technique might be useful in producing ink at a large scale. A similar effort was made to print counter electrodes of graphene in a dye-sensitized solar cell[45]. However, the $MoS_2$ nanosheets are semiconducting and expected to find applications in printing electrodes of capacitors and batteries, inkjet devices, and flexible and wearable electronic devices.

**Conclusions**

To summarize, we exfoliated the $MoS_2$ bulk crystals in DI water by a novel HP LPE technique resulting in 4-7 layered nanosheets. The Tyndall effect confirmed that the dispersion is colloidal. The UV-Visible and Raman spectroscopy results confirmed the successful exfoliation of the material. TEM and SEM images indicated the size of the nanosheets to be a few hundred nanometers. Among the surfactants SDBS, SC, and TBAB, SDBS was found to be better suited for producing high yield and more stable $MoS_2$ nanosheets. Increased soaking duration is found to help obtain thinner nanosheets with a higher yield. The surfactant was efficiently removed by washing the nanosheets with acetone and isopropanol during vacuum filtration. The nanosheets' final concentration was 1.45 mg/ml when we started with an initial concentration of 20 mg/ml, thus taking the yield to be as high as 7.25%. The parameters such as exfoliation cycles, applied pressure, and surfactant concentration can be used to regulate the nanosheets' concentration and lateral size. MoS$_2$ nanosheets were active for the HER with lower Tafel slopes of 189 mV dec$^{-1}$ and 223 mV dec$^{-1}$ (for 60 and 15 days nanosheets) than the bulk (263 mV dec$^{-1}$). We were able to achieve a very high hydrogen evolution rate of 30.13 mmol g$^{-1}$ h$^{-1}$ under ambient laboratory conditions. Thus, high-pressure exfoliation is an industrial scale, eco-friendly and tunable technique to produce highly pure nanosheets which are suitable for hydrogen production.

# SUPPORTING INFORMATION

**Section S1: Materials**

MoS$_2$ bulk powder (Asbury Graphite Mills, Inc.), Sodium dodecylbenzene sulfonate (SDBS) from SRL, Sodium cholate (SC) from Sigma Aldrich, Tetrabutyl Ammonium bromide (TBAB) from Loba Chemie, and distilled water (directly collected for the Millipore water purification system of conductivity 18.2 MΩ) were utilized for the study. All reagents were used as such without any further purification. AAO and PVDF substrates (pore size ~ 0.1 μm for both) were purchased from Merck. Sylgard 184 Silicone elastomer (PDMS prepolymer) and curing agent were used for preparing in-plane ion transport devices.

**Section S2: Physical characterization**

The crystalline structure of prepared MoS$_2$ nanosheets was determined by X-ray diffraction studies using an automated multipurpose X-ray diffractometer by Rigaku SmartLab at 1.54 Å Cu Kα. Raman spectroscopic measurements were performed using LabRAM HR Evolution (Horiba Scientific) Raman spectrometer equipped with 532 nm laser source and 100X long working distance microscope objective. The absorption studies of MoS$_2$ nanosheets were evaluated using a UV–visible spectrophotometer (Malvern). The cuvette path length was 1 cm. Surface morphology and elemental analysis of the MoS$_2$ nanosheets were done using FE-SEM, JSM7600F with Oxford energy dispersive X-ray spectroscopy (EDX) attachment. Transmission electron microscope studies were done using high-resolution TEM, TF-Themis 300 to determine the morphology and crystallinity of prepared MoS$_2$ nanosheets.

**Section S3: Sample preparations for characterizations**

TEM sample was prepared on an ultrathin C film on lacey Carbon (Support film, 400 mesh, Cu). Raman samples were prepared on a Si/SiO$_2$ wafer by drop-casting nanosheets cleaned from surfactant. The sample for SEM was prepared by drop-casting the surfactant-washed nanosheets (by the method described in the main text) on a Si wafer. The sample was coated with Pt before the experiment. UV-visible absorbance experiment was carried out using a 1 cm path length cuvette. The solutions were diluted from the master solution, which was the solution obtained after centrifugation. Samples for Zeta potential measurement were prepared by diluting the master solution with DI-water in the ratio of 1:10.

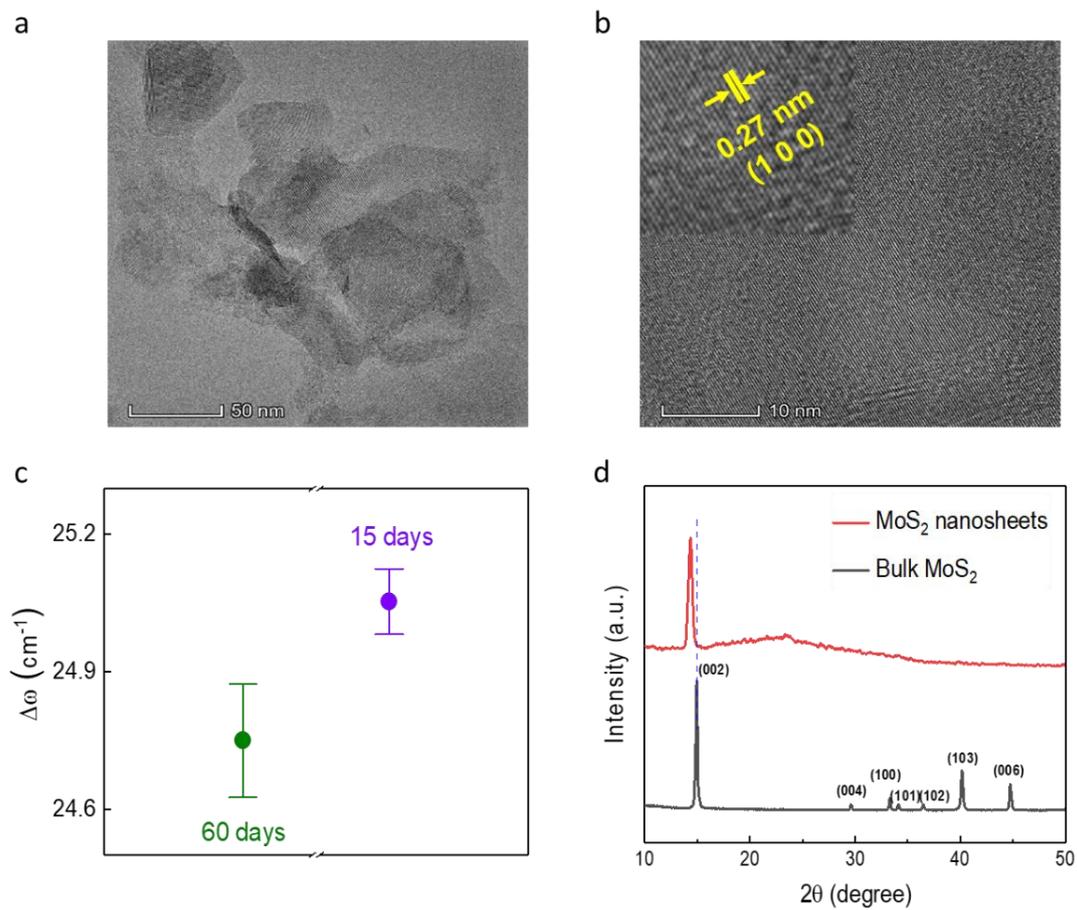

**Fig. S1:** (a) HR-TEM images of MoS$_2$ nanosheets obtained from the high-pressure LPE technique. (b) The interlayer spacing of 0.27 nm represents the (1 0 0) plane in the exfoliated nanosheets. (c) Raman shift for MoS$_2$ nanosheets soaked for 15 days is higher than 60 days. (d) XRD patterns of bulk MoS$_2$ powder and 15 days nanosheets.

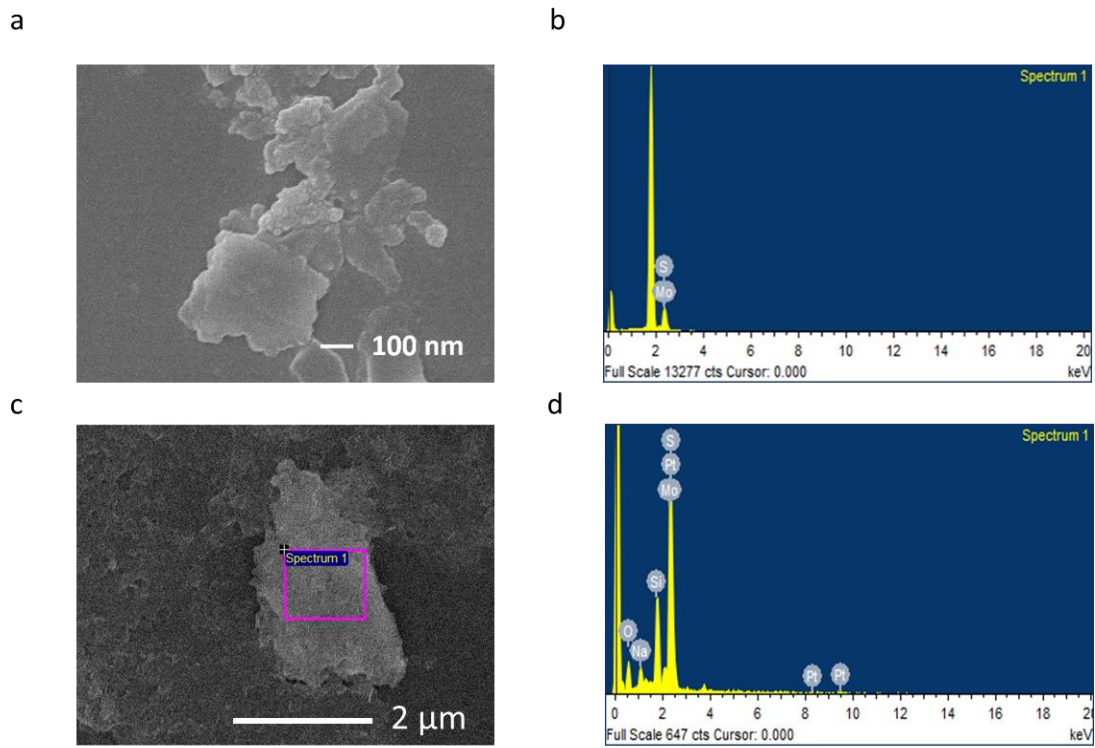

**Fig. S2:** (a) SEM image and (b) the corresponding EDS spectra of MoS$_2$ nanosheets that is washed with acetone and isopropanol to remove the surfactants. (c) SEM image and (d) the corresponding EDS spectra of MoS$_2$ nanosheets before washing, showing the presence of surfactant (in the form of Na, from SDBS).

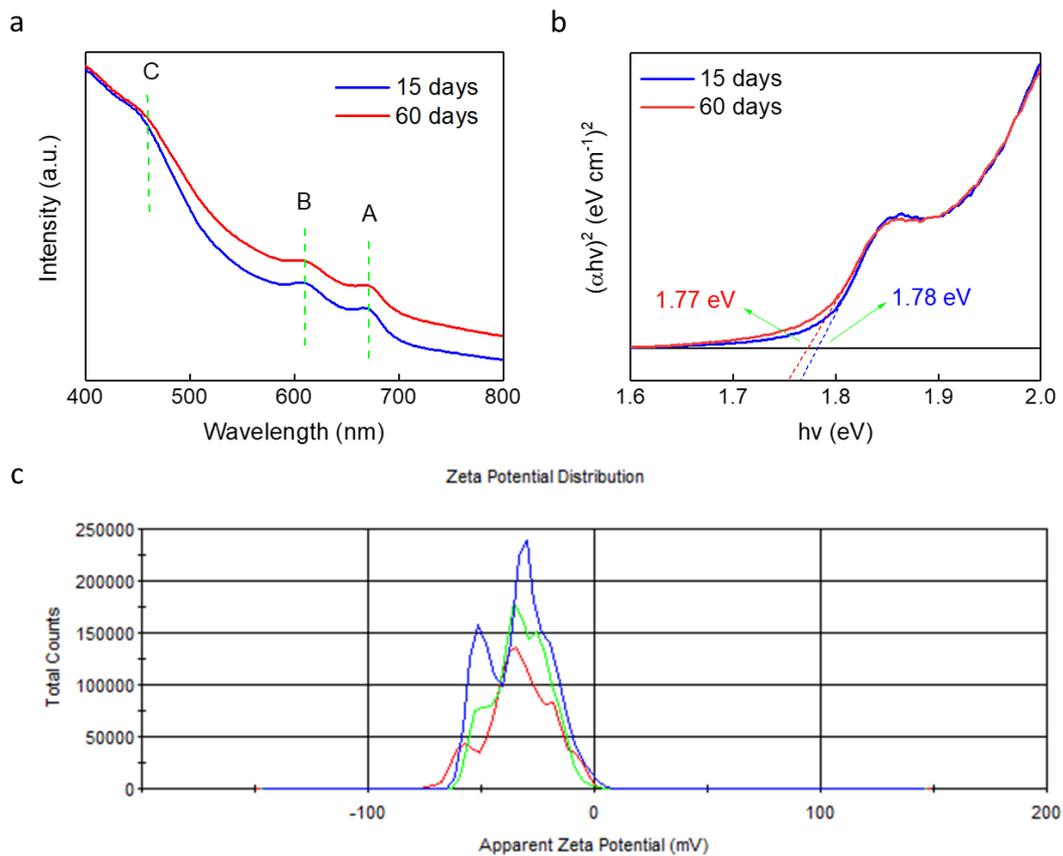

**Fig. S3:** (a) UV-Visible spectrum and (b) bandgap of MoS$_2$ nanosheets that were soaked in SDBS surfactant for different soaking durations. (c) Zeta Potential measurements provided a mean value of -33 mV. The different curves are from three consecutive measurements.

**Section S4: Influence of critical parameters**

*S4.1: Influence of cycles on the concentration and lateral size:*

It is necessary to understand how the cycles affect the final concentration. Fig. S4 (a) and (b) show the effect of varying the number of cycles in the procedure on the concentration of MoS$_2$ nanosheets as well as the lateral size of the end product. It is clear from Fig. S4 (a) that with increasing the number of cycles at a pressure of 50 bar, the concentration also increases. The same trend is observed for higher pressures of 100 bar and 150 bar. It is found that at higher pressures, we obtain higher concentrations. It could be due to the fact that as the number of cycles increases, the unexfoliated flakes also experience shear while the exfoliated flakes experience shear, again and again, eventually increasing the concentration. As expected, the lateral size of MoS$_2$ nanosheets decreases when increasing the number of cycles at a pressure of 50 bar, as shown in Fig. S4 (b). However, at higher pressures 100 bar and 150 bar, the lateral size remains unchanged with respect to the number of cycles. This could be because the nozzle size is 279.4 µm, so there is a limit to the maximum fragmentation that can be achieved.

*S4.2: Influence of applied pressure on the concentration and lateral size:*

The study of the influence of applied pressure is necessary to understand its role with the variation of other parameters such as cycle number. Fig. S4 (c) and (d) show the influence of applied pressure on the concentration and the lateral size of the MoS$_2$ nanosheets. Fig. S4 (c) shows that at a fixed cycle, the concentration increases as pressure increases. For a higher cycle number, the concentration is higher than that for a lower cycle number. It could be because the applied force on each flake per unit area increases. Fig. S4 (d) shows the influence of applied pressure on the lateral size. The lateral size decreases with the increase in pressure, as expected. This effect is more pronounced for lower cycle numbers than for higher cycle numbers. As the size of the flake decreases with the increase in the number of cycles (Fig. S4 (b)), its area decreases, and now it will experience less force (to maintain the same applied pressure) in higher cycles. Therefore, the higher cycle has a less prominent effect on the increase in applied pressure than the lower cycle, as shown in Fig. S4 (d).

*S4.3: Influence of surfactant concentration on the concentration and lateral size:*

Fig. S4 (e) and (f) show the concentration of the MoS$_2$ nanosheets with different surfactant concentrations at pressures of 50 bar and 100 bar, respectively. The nanosheets concentration shows an increase with pressure for a constant surfactant concentration. However, the increase in surfactant concentration slightly reduces the nanosheets concentration at 50 bar and follows the same trend for a different number of cycles. It is probably because the flakes experiencing the shear stress is less due to larger surfactant molecules leading to a lower nanosheets concentration. A larger pressure helps to expose the flakes to shear stress and thus increases the nanosheets' concentration. However, at large surfactant concentrations, even high pressure is not helpful as the destabilization of the surfactants from the layers is dominated.

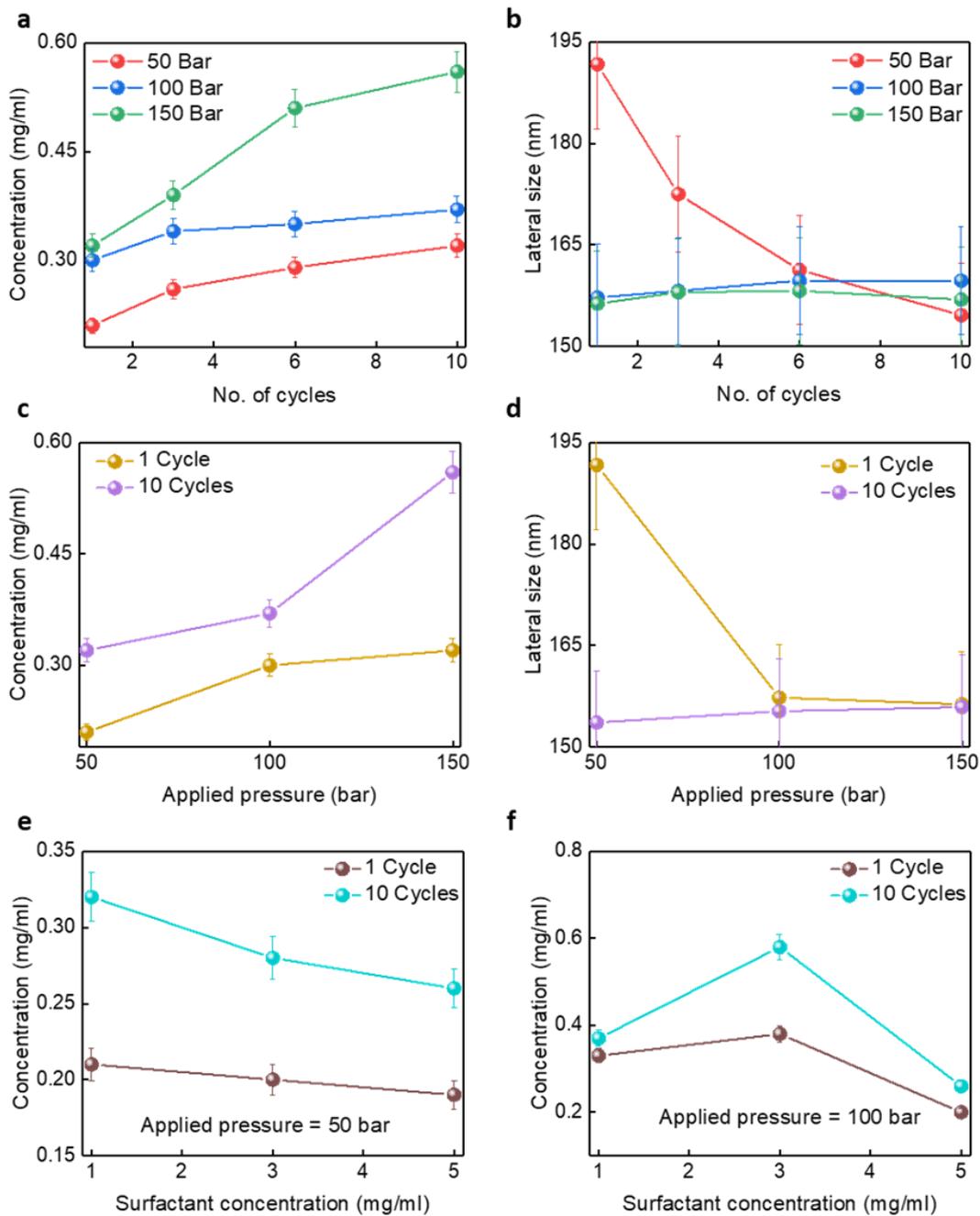

**Fig. S4:** Influence of critical parameters on final concentration and lateral size of the MoS$_2$ nanosheets. The error bar indicates the standard deviation.

**Section S5: Calculations**

*S5.1: Estimation of Crystallize size from XRD measurement:*

Scherrer formula was used to estimate the average crystallite size:

$$d = \frac{0.9\, \lambda}{\beta \cos \theta}$$

λ = wavelength of the X-rays = 1.54 Å,
2θ = peak position, θ = 14.29°,
β = FWHM at θ = 0.74° = 0.0129 radians

⇨  d = 10.84 nm

*S5.2: Estimation of concentration from UV-Visible spectroscopy:*

Lambert-Beer law is A/l=αC, where A is the absorbance, l is cell length (1 cm cuvette), α is the extinction coefficient, and C is the concentration.

At 610 nm, A=0.099, α = 0.10 ml mg$^{-1}$ cm$^{-1}$

⇨  C = 0.95 mg/ml

At 670 nm, A=0.077, α = 0.08 ml mg$^{-1}$ cm$^{-1}$

⇨  C = 0.96 mg/ml

*S5.3: Estimation of hydrogen evolution rate:*

From our calibrated gas chromatography system, an area of 634224.5 mV min corresponds to 8.17 x 10$^{-7}$ mol.

Area under the hydrogen peak (Fig. S6 (b)) = 17518 mV min.
No. of mol of hydrogen produced = 2.26 x 10$^{-8}$ mol

200 ml slurry contains 25 mg nanosheets.
⇨  3 µl slurry contains 37.5 x 10$^{-8}$ g nanosheets

Amount of hydrogen produced = 60.27 mmol g$^{-1}$ in 2 hours
Therefore, hydrogen evolution rate = 30.13 mmol g$^{-1}$ h$^{-1}$

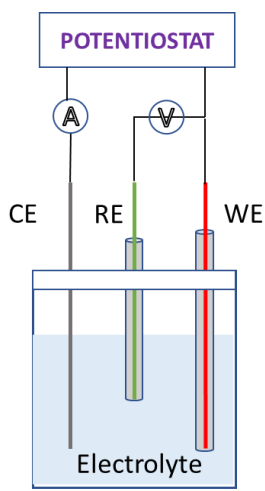

**Fig. S5:** The schematic of the three-electrode system for electrochemical studies

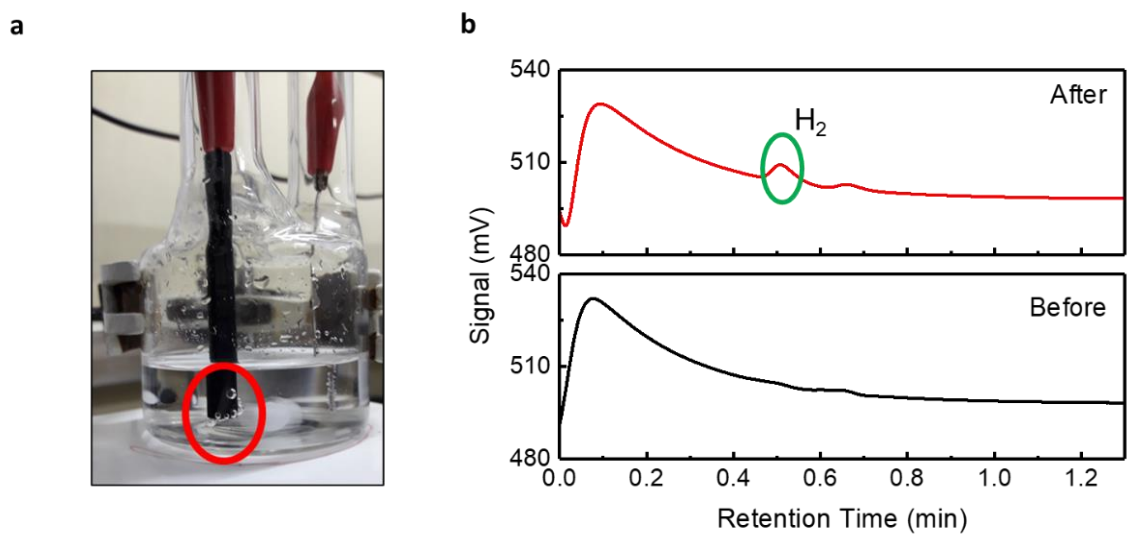

**Fig. S6:** (a) Bubbles observed at the WE (60 days MoS$_2$ nanosheets-coated glassy carbon electrode) during chronoamperometry. (b) Gas chromatogram of 60 days-soaked sample before and after chronoamperometry. The peak at 0.5 min retention time is visible in the chromatogram obtained after chronoamperometry, showing the evolution of hydrogen.

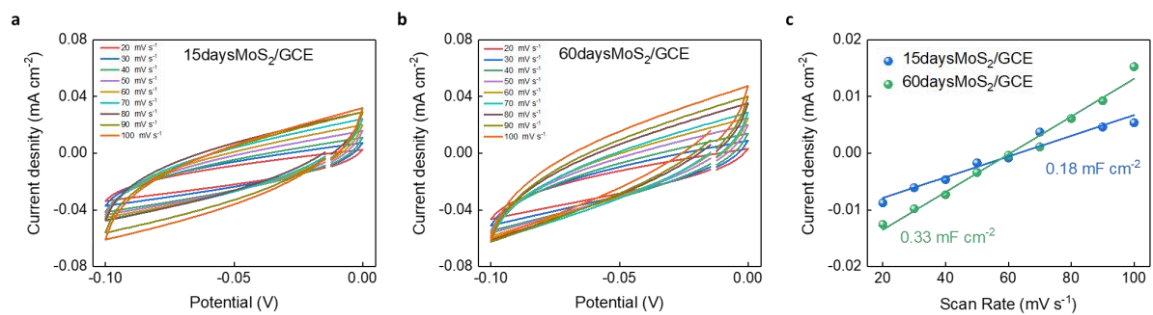

**Fig. S7:** CV curves of (a) 15 days and (b) 60 days nanosheets at a narrow voltage window from 0 V to -0.1 V at different scan rates ranging from 20-100 mV s$^{-1}$. (c) The capacitive currents at -0.05 V as a function of scan rate for 15 and 60 days nanosheets. The 60 days nanosheets have a larger $C_{dl}$ value, indicating their higher surface area than the 15 days nanosheets.

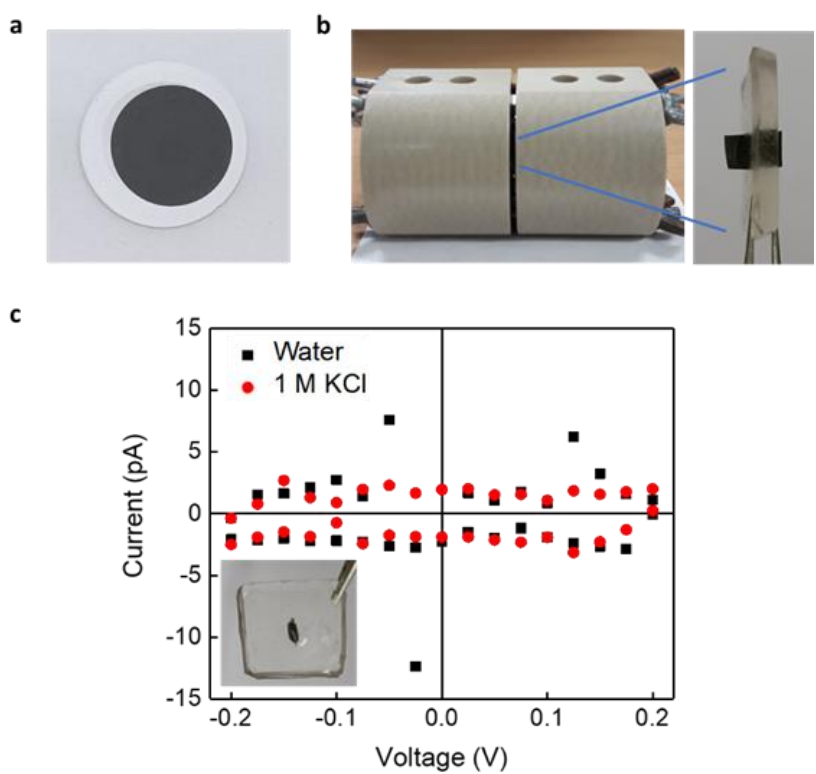

**Fig. S8:** (a) MoS$_2$ membrane prepared through vacuum filtration on PVDF of pore size 0.1 μm. (b) The electrochemical setup for ion-transport study. An enlarged view of the membrane device (encapsulated between PDMS with its edges exposed) is shown on the right. (c) Current-voltage (*I-V*) characteristics of the sample. Inset shows the top view of the membrane device shown in (b).